\documentclass[lettersize,journal]{IEEEtran}
\usepackage{amsmath,amsfonts}
\usepackage{amssymb}
\usepackage{algorithmic}
\usepackage{algorithm}
\usepackage{array}
\usepackage[caption=false,font=normalsize,labelfont=sf,textfont=sf]{subfig}
\usepackage{textcomp}
\usepackage{stfloats}
\usepackage{url}
\usepackage{verbatim}
\usepackage{graphicx}
\usepackage{cite}
\usepackage{subfig}
\usepackage{booktabs}
\usepackage{hyperref}
\usepackage{multirow}
\usepackage{float}
\usepackage{authblk}
\usepackage{adjustbox}
\usepackage{wrapfig}

\begin{document}

\title{AudioNet: Supervised Deep Hashing for Retrieval of Similar Audio Events}

\author[1]{Sagar Dutta,~\IEEEmembership{Member,~IEEE,}}     
\author[2]{Vipul Arora, ~\IEEEmembership{Member,~IEEE}}
\affil[1]{RITMO Centre for Interdisciplinary Studies in Rhythm, Time and Motion \protect\\ Department of Musicology, University of Oslo}
\affil[2]{Department of Electrical Engineering, Indian Institute of Technology Kanpur, India}

\maketitle

\begin{abstract}

This work presents a supervised deep hashing method for retrieving similar audio events. The proposed method, named AudioNet, is a deep-learning-based system for efficient hashing and retrieval of similar audio events using an audio example as a query. AudioNet achieves high retrieval performance on multiple standard datasets by generating binary hash codes for similar audio events, setting new benchmarks in the field, and highlighting its efficacy and effectiveness compare to other hashing methods. Through comprehensive experiments on standard datasets, our research represents a pioneering effort in evaluating the retrieval performance of similar audio events. A novel loss function is proposed which incorporates weighted contrastive and weighted pairwise loss along with hashcode balancing to improve the efficiency of audio event retrieval. The method adopts discrete gradient propagation, which allows gradients to be propagated through discrete variables during backpropagation. This enables the network to optimize the discrete hash codes using standard gradient-based optimization algorithms, which are typically used for continuous variables. The proposed method showcases promising retrieval performance, as evidenced by the experimental results, even when dealing with imbalanced datasets. The systematic analysis conducted in this study further supports the significant benefits of the proposed method in retrieval performance across multiple datasets. The findings presented in this work establish a baseline for future studies on the efficient retrieval of similar audio events using deep audio embeddings. Source code is available at: \url{https://github.com/sagar0dutta/AudioNet}\footnote{This work was done at IIT Kanpur, India, and supported by Prasar Bharati, India’s National Broadcasting Company.}

\end{abstract}


\section{Introduction}
\IEEEPARstart{T}{he} rapid growth of audio content on the internet has significantly increased the volume of digital information. This growth necessitates efficient audio search algorithms capable of handling the extensive databases that have become commonplace. Traditional search methods, particularly linear searches, are not feasible for databases comprising billions of audio files due to their prohibitive time and memory requirements.

As audio data production and storage continue to expand, there is a pressing need for effective audio retrieval methods. Although considerable research has been conducted in the field of visual data retrieval, audio retrieval has not received the same level of attention, despite its importance \cite{para2a}. For instance, in the media and entertainment industry, efficient retrieval of similar audio events based on user queries can dramatically refine content recommendation engines \cite{para2b}, offering users personalized audio experiences. Similarly, in educational contexts, particularly in language learning platforms, the ability to retrieve and compare audio samples can significantly enhance pronunciation tools \cite{para2c}, providing learners with immediate feedback and examples from native speakers. These applications not only demonstrate the broad utility of advanced audio retrieval systems but also underscore the need to develop more sophisticated methods tailored to the specific challenges and opportunities presented by audio data \cite{para2d}.

This study examines various hashing methods for audio data retrieval, categorized into data-independent and data-dependent approaches. Data-independent methods, such as fingerprinting \cite{wang2003industrial} and locality-sensitive hashing \cite{para3a}, offer robustness to data variations without the need for model training. However, their reliance on longer hash codes for achieving high precision necessitates increased storage space and compromises search recall efficiency. Conversely, data-dependent hashing methods \cite{singh2023simultaneously} train similarity hash functions using datasets to maintain similarity measures from the original feature space in the Hamming space. These methods are further divided into unsupervised, semi-supervised, and supervised techniques based on the availability of labeled data during the training phase. Despite the advancements in deep hashing methods within the visual domain, traditional convolutional neural network (CNN) models often face challenges when applied to imbalanced datasets, a common issue in the audio domain \cite{para3c}.

This study introduces AudioNet, a deep-learning method for hashing similar audio events for retrieval, representing a notable contribution in the audio retrieval domain. By focusing on the efficient retrieval of similar audio events based on audio queries, AudioNet addresses the gaps identified in previous research, which mainly concentrated on exact match or cross-modal retrieval. The framework is designed to tackle challenges such as ill-posed gradients and data imbalance, employing deep feature learning and binary hash encoding to achieve superior retrieval performance. Comprehensive evaluations on publicly available audio event datasets have tested AudioNet’s effectiveness. By demonstrating its potential to set new benchmarks, AudioNet aims to establish a baseline that will guide the development of more effective audio retrieval systems and foster future advancements in the field. The main contributions of the paper are as follows:

\begin{itemize}

\item Introducing deep audio embeddings to address and formalize the challenge of efficient retrieval of similar audio events using an audio-based query.

\item Presenting an approach that achieves good retrieval performance while accommodating imbalanced datasets for hashcode learning.

\item Proposing a novel loss function by incorporating weighted contrastive loss and weighted pairwise loss with hashcode balancing for efficient audio events retrieval.

\item Conducting systematic analysis and demonstrating significant performance benefits of the proposed method in audio event retrieval across multiple datasets.
\end{itemize}

\section{Related Work}

Learning to hash methods can be classified into unsupervised, semi-supervised, and supervised approaches, depending on the availability of labeled data during the training process. Unsupervised hashing techniques utilize unlabeled data to develop hash functions that effectively preserve the proximity between neighboring instances. Iterative Quantization \cite{d26} iteratively adjusts continuous values to binary to maintain the data structure, while Spectral Hashing \cite{d27} solves a spectral graph partitioning problem to ensure similar data points receive similar codes. Locality-sensitive hashing \cite{para3a} hashes items such that similar items map to the same buckets with high probability, Anchor Graph Hashing \cite{agh} uses anchors to summarize data space and hashes points based on their relationship with these anchors, and Product Quantization \cite{pq} decomposes the space into a Cartesian product of low-dimensional subspaces for efficient similarity search and storage. These are examples of unsupervised methods. Semi-supervised and supervised techniques incorporate label information into the learning process to improve the quality of hashing. Semi-Supervised Hashing \cite{d35} leverages both labeled and unlabeled data, Kernel-based Supervised Hashing \cite{d36} uses a kernel function to transform data into a more separable feature space, Minimal Loss Hashing \cite{d28} focuses on minimizing the loss between original and Hamming distances, and Supervised Discrete Hashing with Point-wise Labels \cite{d37} learns binary hash codes for individual data points based on their labels. Data-dependent methods, unlike data-independent methods, allow for compact coding, making them more practical for real-world applications.

Recent advancements in deep hashing within the image domain have introduced efficient models that contribute significantly to binary code generation and data retrieval systems. Hashing methods such as Deep Supervised Discrete Hashing (DSDH) \cite{dsdh}, Greedy Hash \cite{greedy}, Deep Hashing Network (DHN) \cite{dhn}, Deep Pairwise-Supervised Hashing (DPSH) \cite{dpsh}, and OrthoHash \cite{hoe2021one} have demonstrated promising results in image retrieval tasks achieving mean average precision above 75\%. The use of binary hash codes has significantly reduced retrieval time and memory usage. Conventional CNNs, successful in image classification, show promise as feature extractors for hashing problems due to their ability to capture semantic information. However, CNNs struggle with imbalanced datasets, which hampers feature generation. Optimizing hash code length is also vital for balancing precision and retrieval efficiency, as explored by He et al. \cite{he2016deep}. Thus, there is a need to balance hash code length with retrieval efficiency in designing effective audio retrieval systems.


Audio event classification is a rapidly growing field of research that is becoming increasingly popular due to its wide range of applications in areas such as accessibility devices, health monitoring, audio content understanding, and surveillance. A pre-defined ontology \cite{a01} is followed to annotate the audio events, where the annotations are used for supervised learning. Despite several attempts to learn more accurate embeddings of audio events, not much work has been done to facilitate large-scale efficient audio retrieval based on the audio query. For example, Koepke et al. \cite{a3} investigated cross-modal retrieval tasks involving text and audio data, to retrieve audio content based on written descriptions, as well as retrieving text descriptions based on audio content. An algorithm for similarity search is proposed \cite{a4} for animal sound recordings in large archives. The authors present a score-audio music retrieval system in \cite{a6} to retrieve audio based on music notation. Furthermore, deep learning to hash methods has been also proposed for audio retrieval. For example, Ye et al. \cite{a9} proposed a system to detect cover songs using supervised deep hashing. An unsupervised hashing method is proposed by Panyapanuwat et al. \cite{a8} for content-based audio retrieval using a deep neural network. Tran et al. \cite{a7} propose a deep hashing algorithm for speaker identification and retrieval that considers speaker individuality by using auditory sparse representations. Furthermore, Jati et al.\cite{a10} proposed a deep hashing framework to transform weak audio embeddings into low-dimensional hash codes for efficient retrieval of audio events. Recent advancements in this field include Kim et al.'s \cite{kim2022boosted} development of Boosted Locality Sensitive Hashing for source separation, introducing an adaptive boosting approach to learn discriminative binary hash codes for single-channel speech denoising. 
Moreover, hashing methods have been developed in video retrieval to facilitate efficient and scalable retrieval using video as the query. Li et al. introduced self-supervised video hashing using bidirectional transformers, which overcomes the limitations of traditional unidirectional models by fully exploiting the bidirectional correlations between video frames \cite{vid01}. Additionally, Li et al. proposed an unsupervised variational video hashing (UVVH) method that addresses the shortcomings of RNN-based approaches, which often struggle with content forgetting and fail to capture global information. The UVVH method employs a 1D-CNN-LSTM model to process long frame sequences in a parallel and hierarchical manner, ensuring the extraction of salient and global features, thereby producing reliable binary codes for video retrieval \cite{vid02}.  

There has been little research on the retrieval of similar audio events based on audio queries. Audio fingerprinting systems \cite{singh2023simultaneously} retrieve general audio that exactly matches the user's search query. Zhang \emph{et al.} \cite{zhang2018siamese} search audio by vocal imitation. Ottomechanic \cite{morrison2019otomechanic} uses automobile engine sound recordings to search for similar audio in a database for diagnosing engine faults. 
The discussed studies primarily focus on retrieving audio events through exact matching or identifying the original sound, using text queries for cross-modal retrieval, and employing vocal imitations. However, these methods have limitations. Textual descriptions often fail to capture the audio content in its entirety, and vocal imitations fall short in replicating authentic sound features. This highlights a noticeable gap in research concerning the effective retrieval of similar audio events through audio-based queries. Furthermore, there is a need to explore this within the framework of unbalanced datasets and to assess the efficacy of optimized hash codes applied across various datasets.

This work introduces an efficient system for retrieving similar audio events based on an audio query, which involves accurately and quickly searching and retrieving similar audio events from a large audio database. The study also aims to establish benchmark performance metrics that can be used for future research in this area. A comprehensive evaluation of the proposed method is conducted using publicly available audio event datasets. By evaluating the performance of the proposed method against established standard datasets, the study aims to provide insights and guidance for further research and development in the field of similar audio event retrieval using an audio query. 

\section{Hashing Network and features}

\begin{figure*}[t]
\centering
\includegraphics[width=0.8\textwidth]{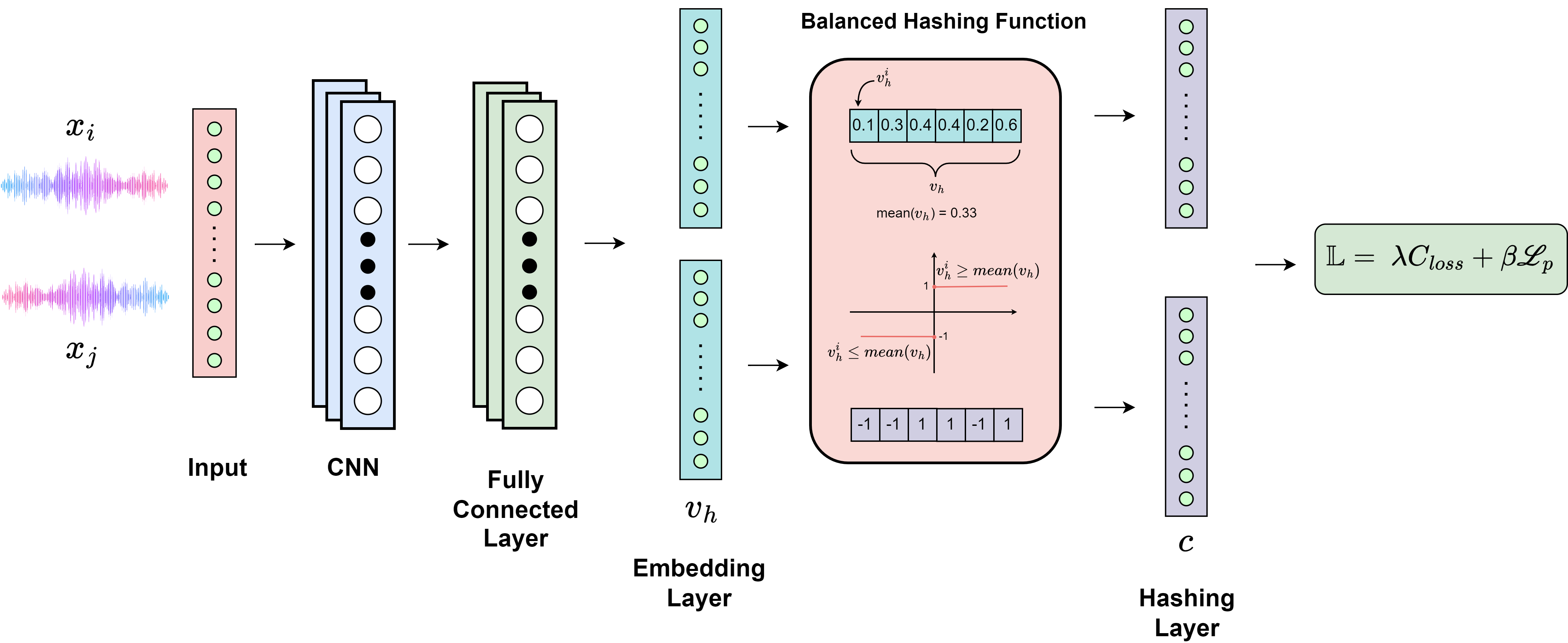}
\caption{Overview of the end-to-end architecture of our proposed method. Here the CNN learns to extract meaningful features from the audio events, the fully connected hash layer projects these features into a lower-dimensional hash representation, the balanced binarization function converts the continuous representations into binary codes, and the weighted contrastive loss is used as a training objective to ensure that the similarity between audio events is preserved in the learned hashcodes.}
\label{main}
\end{figure*}

Large datasets and sophisticated models have significantly advanced visual and audio data processing, enhancing classification and retrieval tasks. In the visual domain, ImageNet and Nus-wide, along with frameworks like ResNet \cite{m4}, AlexNet \cite{m1}, and VGG \cite{m2}, have been instrumental in improving image classification performance. Similarly, in audio, pre-trained models such as YAMNet \cite{howard2017mobilenets} and Pretrained Audio Neural Networks (PANNs) \cite{kong2020panns} have excelled in tasks like audio pattern recognition, with EfficientNetV2 \cite{nguyen2023fruit} also being adaptable for audio. These advancements underline the importance of advanced feature extraction for audio retrieval system enhancement. This study evaluates the potential benefits of similar-sized datasets and frameworks in audio event retrieval.

The AudioNet architecture processes pairs of audio inputs through a sequence that includes deep representation learning and binary hash encoding. This sequence encompasses: (1) utilizing a convolutional neural network (CNN) to learn complex representations, (2) employing a dense hash layer to convert these complex representations into a K-dimensional space, (3) applying a balanced sign activation function to translate the K-dimensional representations into K-bit binary hash codes, and (4) integrating a novel loss function tailored for maintaining similarity in learning processes, especially when dealing with data imbalances. We chose Mel-frequency cepstral coefficients (MFCCs) in our study for their ability to provide a compact representation of audio data, leading to computational efficiency. This choice is practical, particularly given our computational limitations and the necessity for resilience against noise and variations in recording conditions found in the dataset. The reduced computational requirements of MFCCs render them a sensible option for applications in real-world settings. An audio event can be thought of as a succession of frames $F = F_1, F_2,..., F_n$, where $F_i$ is an audio frame described by its features. The Mel-frequency cepstral coefficients are extracted with a window length of 4096 and a hop length of 1024 on a sampling rate of 44,100Hz. This equates to $\sim$92ms window with $\sim$23ms overlap, and 40 resulting coefficients.

\section{Approach}

Given a dataset of \( N \) audio events denoted as \( A = \{a_i\}_{i=1}^N \), along with their associated label information \( L = \{l_i\}_{i=1}^N \) where \( l_i \) is a one-hot encoded vector of dimension \( D \), with \( D \) representing the number of unique labels. Each audio event \( a_i \) is represented by an \( M \)-dimensional feature vector \( x_i \in \mathbb{R}^M \). For a pair of audio events \( a_i \) and \( a_j \), we denote their binary hash codes as \( c_i \) and \( c_j \), respectively, where \( c_{i} \in \mathcal{C} \). Each pair of audio events is also associated with a similarity label \( s_{ij} \in S \), where \( S = L\cdot L^{\intercal} \).

The similarity label \( s_{ij} = 1 \) indicates that the pair of audio events \( a_i \) and \( a_j \) belong to the same class, and \( s_{ij} = 0 \) otherwise. The hashing network learns a mapping function \( f: \mathbb{R}^M \rightarrow \mathcal{C} \), where \( \mathcal{C} \) represents a set of hash codes of \( k \) bits. This mapping function \( f \) transforms the feature vectors \( x_i \) from the input space \( \mathbb{R}^M \) to the Hamming space \( \mathcal{C} \), ensuring that the pairwise similarity between audio events is preserved in the compact binary hash codes.

\begin{figure*}[t]
\centering
\includegraphics[width=0.8\textwidth]{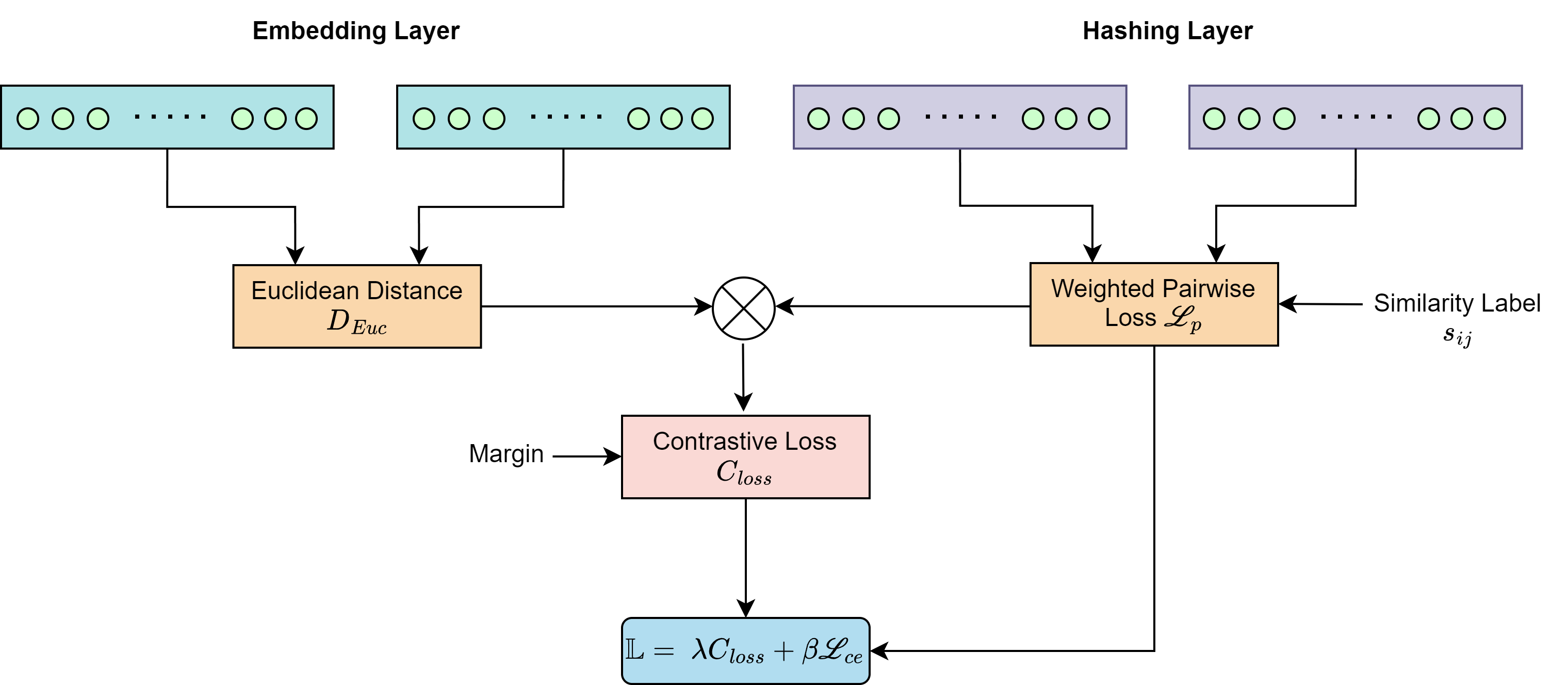}
\caption{Overview of the computation of the proposed loss}
\label{Loss}
\end{figure*}

The proposed model in this study processes pairs of audio events \( \{(a_i, a_j, s_{ij})\} \) and includes several components: a convolutional neural network (CNN) that learns deep representations for each feature vector \( x_i \), a fully-connected hash layer that maps these deep representations to a \( K \)-dimensional representation \( v_h = \{ v_{h}^{i}\}_{i=1}^{K} \), \( v_h \in \mathbb{R}^K \), a balanced binarization function that converts the deep representations into balanced binary hash codes \( c \), and a novel weighted contrastive loss and weighted pairwise loss to maintain the similarity-preserving properties. The architecture of the proposed model is shown in Fig. \ref{main}. The CNN learns to extract meaningful features from the audio events, the fully-connected hash layer projects these features into a lower-dimensional hash representation, the balanced binarization function converts the continuous representations into binary codes, and the proposed weighted loss is used as a training objective to ensure that the similarity between audio events is preserved in the learned hashcodes. Fig. \ref{Loss} depicts the computation of the proposed loss. The combination of the above components produces an effective audio hashing model that can generate compact binary codes for efficient audio retrieval and similarity-based tasks.

The pairwise loss hashing method relies on the similarities between two feature samples to generate a hash. The method's performance benefits hinge on the availability of a strong paired construction. There are several factors that need to be managed, including the ratio of samples with similar characteristics to those with different characteristics and the incorporation of multi-class samples into the overall distribution. It is challenging to construct balanced pairwise samples due to issues such as imbalanced interclass samples, relatively small variances across classes, and limited access to labeled data. In this study, we find a way around this problem by enhancing conventional contrastive loss with label data, allowing us to construct discrete hash codes. The traditional way to formulate the contrastive loss \cite{hadsell2006dimensionality} for a pair of audio samples is,

\begin{equation}
\label{CL}
\resizebox{.9\hsize}{!}{$C_{loss} = \frac{1}{2} S \left\|v_{h}^i- v_{h}^j \right\|_{2}^2 + \frac{1}{2} (1-S)\max(0, p-\left\|v_{h}^i- v_{h}^j \right\|_{2}^2)$}
\end{equation}

where $p$ is the margin parameter set to 1 and $\left\|v_{h}^i- v_{h}^j \right\|_{2}^2$ is the distance function. The margin value is utilized to control the separation between positive and negative samples. Furthermore, most convolutional neural network architectures are not designed to be updated with discrete (binary) outputs. As a result, the distance is computed in Euclidean space.

The proposed loss improves hashcode intraclass difference and interclass distance, as compared to the traditional contrastive loss. Additionally, the contrastive loss incorporates pairwise similarity information with the Euclidean distance, thereby reducing differences between similar hashcodes. The weighted contrastive loss is computed based on both the weighted pairwise similarity of hashcodes and the label information. Now for a given pair of hashcodes, denoted as $c_i$ and $c_j$, with a similarity label of $s_{ij}$,

\begin{align}
\label{proba}
P(s_{ij}|c_i, c_j) &= \phi(c_i^{T}\cdot c_j)^{s_{ij}}(1-\phi(c_i^{T}\cdot c_j))^{1-s_{ij}}
\end{align}

where $\phi(z)=1 / (1+e^{-\alpha z}) $ is the sigmoid fuction, and the ${c_i^{T}\cdot c_j}$ is the inner product of hash codes ${c_i}$ and ${c_j}$. For binary hash codes, the inner product, which counts matching versus non-matching bits, offers a more relevant similarity measure. It accurately indicates the number of identical bits in two hash codes, a key factor in binary data comparison. Conversely, euclidean distance is less suitable for binary data, where differences are absolute and not gradual, making it less effective and intuitive for capturing similarity in high-dimensional binary spaces. Equation \ref{proba} makes it clear that higher the $P(s_{ij}|{c_i, c_j)}$, the larger the inner product between ${c_i}$ and ${c_j}$. As a result, ${c_i}$ and ${c_j}$ can be classified as similar. The Weighted Maximum Likelihood (WML) estimation is given by,

\begin{align}
\label{pairwiseloss}
    \log P(S|\mathcal{C}) &= \sum_{s_{ij}}w_{ij}\, \log P(s_{ij}|c_i, c_j)
\end{align}

From Eq. \ref{proba} and Eq. \ref{pairwiseloss}, we get

\begin{align}
\label{Dce}
    \mathcal{L}_{p} &= \sum_{s_{ij}}w_{ij}\left(\log (1+e^{\alpha c_i^{T}\cdot c_j})-\alpha s_{ij}c_i^{T}\cdot c_j\right)
\end{align}

where,

\begin{align}
\label{weight}
    w_{ij} = \log \left ( \frac{N_{similar} + N_{dissimilar}}{s_{ij} \cdot N_{similar} + (1 - s_{ij}) \cdot N_{dissimilar}} \right )
\end{align}



In retrieval systems, it is commonly observed that the availability of similarity information is quite limited. Specifically, the number of pairs exhibiting similarity is significantly lower compared to the number of pairs displaying dissimilarity. As a consequence, an inherent data imbalance arises. To address this issue, a novel weighting mechanism denoted as $w_{ij}$ is introduced for each training pair. This approach aims to address the issue of imbalanced data by assigning weights to the audio event pairs based on the similarity of each specific pair. $N_{similar}$ and $N_{dissimilar}$ are the total number of positive and negative pairs of audio events. Now the new distance in the contrastive loss function is given by,

\begin{equation}
\label{dtotal}
    D_{tot} =  D_{Euc} + \mathcal{L}_{p}
\end{equation}

And Eq. \ref{CL} is revised and stated as Eq. \ref{cl1} given this new distance,

\begin{equation}
\label{cl1}
\resizebox{.7\hsize}{!}{$C_{loss} = \frac{1}{2} S\cdot D_{tot} + \frac{1}{2} (1-S)\cdot \max(0, p-D_{tot})$}
\end{equation}
Here, $p$ is the margin parameter ($=1$ here).
The second improvement is the impact of pairwise audio event similarity on maximizing distances between classes. The hashcode of similar audio pairs must move away from dissimilar pairs. $\mathcal{L}_{p}$, which specifies the weighted pairwise loss between two audio events, is added to the CL to obtain the final loss function as,

\begin{equation}
    \mathbb{L} =  \lambda C_{loss} + \beta \mathcal{L}_{p}
\end{equation}

where $\lambda$ and $\beta$ are the weights.

\subsection{Balanced hashing function}

The efficacy of a hash code not only depends on its length but also critically on the balance of its bit composition. The hash code needs to have an equal distribution of ones and negative ones across different lengths, such as 16, 32, 64, and 128 bits, to ensure each bit effectively contributes to the code's overall representation. This balanced approach is crucial for maintaining the integrity of data features within a given encoding length, regardless of the hash code's size.

Our balanced hashing function is designed with a fixed coding length, emphasizing the distribution of bits within that fixed length. This strategy is crucial for ensuring a balanced representation of data features. The size of the hash code, from a 16-bit to a 128-bit length, directly influences its capacity to accurately represent and differentiate data. The relationship between coding length and retrieval accuracy underscores the rationale behind our choice to maintain a fixed encoding length. Opting for a dynamic encoding length could introduce computational and algorithmic complexities, especially in preserving the balanced distribution of hash values across various lengths. Our method prioritizes balancing computational efficiency with performance, focusing on optimizing the distribution of ones and negative ones within the predetermined encoding length to enhance retrieval accuracy without altering the length itself.

The dynamic thresholding method described in Eq. \ref{binary} plays a pivotal role in achieving this balance. By adapting the threshold value for each deep feature vector $v_h$, we ensure that every bit of the hash code is weighted appropriately, contributing to the overall balance and, consequently, the retrieval accuracy. In this work, rather than relying on a fixed threshold value \cite{balanced}, we instead make use of a dynamic one. Each deep feature vector $\mathbf{v_h)}$ is averaged to determine the current vector's threshold value as,


\begin{equation}
\operatorname{sign}(v_h) =
\begin{cases}
  1, & v_h^i \geq \operatorname{mean}(v_h) \\
 -1, & v_h^i < \operatorname{mean}(v_h)
\end{cases}
\label{binary}
\end{equation}

In Eq. \ref{binary}, the sign function is applied point-wise to each component of the deep feature vector ${v_h}$. For each element ${v_{h}^i}$ of the vector, we compare it against the mean of the entire vector ${v_h}$. If ${v_{h}^i}$ is greater than or equal to this mean, it is assigned a value of 1; otherwise, it is assigned a value of -1. This operation is performed independently for each element $i$ of the vector, ensuring that the resulting hash code reflects the binary decision for each specific feature in the vector. This approach enables us to maintain a balanced and efficient hash code for each specified bit length.

\subsection{Discrete gradient propagation}

The literature extensively supports the notion that continuous relaxation, a technique where an optimization problem is simplified by allowing variables that are naturally discrete to take on continuous values, inevitably leads to quantization errors. This approach is often employed in hashing methods to facilitate gradient-based optimization, as the sign function used to generate discrete hash codes is not differentiable. Traditional methods resort to continuous relaxation due to the ability of the sign function to generate discrete values. Nevertheless, the derivative of the sign function introduces a challenge with the prevalence of zero values. Due to this derived property, back-propagation cannot be used. The traditional continuous relaxation approach addresses this issue but at the cost of some retrieval precision. Previously, this issue has been addressed by incorporating the term regulating quantization error into the loss function. Even though it enhances retrieval efficiency, it does not address the fundamental problem.

Typically, a sign function is used to construct a hash code, denoted as $c = sign(u)$, where $u$ is a real-valued variable. During our training process, we automatically generate discrete hash codes. In the back-propagation phase, we utilize a hard $\tanh$ activation function, which is a piece-wise linear activation function, and employ the straight-through estimator to propagate gradients. If we know the gradient of $\mathbb{L}$ with respect to c and the loss function $\mathbb{L}$, we may calculate the derivative of $\mathbb{L}$ with u, or the gradient of u as,

\begin{equation}
    \Delta_{u} = \frac{\partial \mathbb{L}}{\partial c}\mathbf{1}_{|u|\leq 1}
\end{equation}

The threshold function's (Eq. \ref{binary}) derivative is disregarded by the straight-through estimator, which instead passes the incoming gradient as though it were applied to an identity function. The differentiable function is as follows,

\begin{equation}
    Htan(u) = max(-1, min(1,u))
\end{equation}

In other words, the gradient of u is zero unless the absolute value of u is less than 1, in which case it equals the gradient of c. This is consistent with avoiding continuous relaxation in the hashing approach by allowing discrete values to be introduced directly into the learning process. This method of gradient propagation allows for the generation of discrete hash codes without the customary step of continuous relaxation. 

\section{Experiment}
\subsection{Datasets}

Given the absence of dedicated large-scale audio retrieval datasets, our study utilizes three popular audio event classification datasets to demonstrate the effectiveness of our proposed retrieval approach. These datasets, which include a broad range of environmental sounds, audio events, and acoustic scenes, were selected for their widespread recognition and diversity. This deliberate choice allows us to assess the efficacy of our methodology and provides a foundation for future scalability investigations. The selection encompasses both equally and unequally distributed datasets, offering a comprehensive overview of our approach's applicability to different types of audio classification and retrieval challenges.

The \textbf{Environmental Sound Classification (ESC-50)} dataset\cite{esc} is a collection of 2000 labeled audio recordings. The dataset is loosely arranged into 5 major categories: domestic sounds, animal sounds, natural soundscapes, human sounds, and urban noises. It is organized into equally distributed 50 classes with 5-second audio recordings. The dataset has been partitioned into five folds to enable cross-validation that is comparable, to ensure that segments originating from the same source file are grouped within a single fold.

\textbf{2018 DCASE Task-2} dataset\cite{dcase} contains diverse amount categories such as animal sounds, domestic sounds, musical instruments, nature, etc. The dataset is annotated using Google AudioSet Ontology\cite{audioset}, which is distributed unequally among 41 categories. The dataset comprises 9,500 audio samples. The minimum and maximum number of audio samples in the training set are 94 and 300 respectively. The duration of the audio samples was fixed at 10 seconds.

\textbf{TUT Acoustic Scenes 2017} dataset\cite{tut} includes acoustic scene recordings from 15 different places such as beaches, homes, restaurants, etc. There are 312 samples for every acoustic scene. Each segment from the same original recording was combined into a single subset, which could be either a training dataset or an evaluation dataset.

Furthermore, Table \ref{dataset} presents a comparative analysis of diverse datasets from Dcase datalist \cite{dcaserepo}, both balanced and unbalanced, for testing the robustness of our method. Datasets feature varying durations, sample counts, and classes. Mean average precision (mAP) shows retrieval performance at 64-bit hashing, with higher precision among top-100 retrievals.


\subsection{Evaluation Metric}

The performance of an information retrieval system is assessed by how well the system can retrieve relevant results for a user query. The evaluation metric (order aware metric) that is commonly used in the retrieval literature is the Mean Average Precision (mAP@k), where k is the number of items retrieved. The metric mAP@k shows the performance when only the top K items in the database are looked at. The metric ranges from 0 to 1 and exhibits a high value when positive retrievals, also known as hits, are ranked at the top. Applications like online search engines, where relevant results should be presented first, benefit greatly from having low k values. As k increases, it becomes more likely to contain matches.

The metrics used in this study are given below:

\begin{enumerate}
    \item \textbf{Precision@k} quantifies how many relevant audio events are present in the top-k results. It is given by,
    \begin{equation}
        \textrm{Precision@k} = \frac{\textrm{true\:positives@k}}{\textrm{true\:positive@k + false\:positives@k}}
    \end{equation}

    

    \item \textbf{Mean Average Precision (mAP)} is the mean of average precision for a set of user queries. Given a set of N queries,

    \begin{equation}
        mAP = \frac{1}{N} \sum_{n=1}^{N}\:AP(n)
    \end{equation}
    
    \item \textbf{Precision (Hamming distance $\leq$  2)}  in information retrieval refers to the proportion of relevant documents that are retrieved by a search algorithm when allowing for up to two bits of error in the comparison between the query and the documents.

\end{enumerate}

\begin{table*}[t!]
\caption{Mean Average Precision $\uparrow$ for all retrieved audio events}
\label{table@all}
\centering
\begin{tabular}{|c|cccc|cccc|cccc|}
\hline
\textbf{Method} &
  \multicolumn{4}{c|}{\textbf{ESC-50}} &
  \multicolumn{4}{c|}{\textbf{DCASE}} &
  \multicolumn{4}{c|}{\textbf{TUT}} \\ \hline
 &
  \multicolumn{1}{c|}{\textbf{16b}} &
  \multicolumn{1}{c|}{\textbf{32b}} &
  \multicolumn{1}{c|}{\textbf{64b}} &
  \textbf{128b} &
  \multicolumn{1}{c|}{\textbf{16b}} &
  \multicolumn{1}{c|}{\textbf{32b}} &
  \multicolumn{1}{c|}{\textbf{64b}} &
  \textbf{128b} &
  \multicolumn{1}{c|}{\textbf{16b}} &
  \multicolumn{1}{c|}{\textbf{32b}} &
  \multicolumn{1}{c|}{\textbf{64b}} &
  \textbf{128b} \\ \hline
AudioNet+TCL &
  \multicolumn{1}{c|}{0.295} &
  \multicolumn{1}{c|}{0.397} &
  \multicolumn{1}{c|}{0.439} &
  0.447 &
  \multicolumn{1}{c|}{0.387} &
  \multicolumn{1}{c|}{0.423} &
  \multicolumn{1}{c|}{0.531} &
  0.545 &
  \multicolumn{1}{c|}{0.337} &
  \multicolumn{1}{c|}{0.406} &
  \multicolumn{1}{c|}{0.487} &
  0.503 \\ \hline
AudioNet+WCL &
  \multicolumn{1}{c|}{0.343} &
  \multicolumn{1}{c|}{0.584} &
  \multicolumn{1}{c|}{0.671} &
  0.684 &
  \multicolumn{1}{c|}{0.627} &
  \multicolumn{1}{c|}{0.681} &
  \multicolumn{1}{c|}{0.784} &
  0.793 &
  \multicolumn{1}{c|}{0.499} &
  \multicolumn{1}{c|}{0.654} &
  \multicolumn{1}{c|}{0.754} &
  0.766 \\ \hline
AudioNet+WCL* &
  \multicolumn{1}{c|}{0.376} &
  \multicolumn{1}{c|}{0.632} &
  \multicolumn{1}{c|}{0.692} &
  0.698 &
  \multicolumn{1}{c|}{0.638} &
  \multicolumn{1}{c|}{0.751} &
  \multicolumn{1}{c|}{0.808} &
  0.817 &
  \multicolumn{1}{c|}{0.512} &
  \multicolumn{1}{c|}{0.677} &
  \multicolumn{1}{c|}{0.773} &
  0.780 \\ \hline
\end{tabular}
\end{table*}    

\begin{table*}[t!]
\centering
\caption{Mean average precision $\uparrow$ for top-100 retrieved audio events}
\label{top100}
\begin{tabular}{|l|llll|llll|llll|}
\hline
\textbf{Method} &
  \multicolumn{4}{c|}{\textbf{ESC-50}} &
  \multicolumn{4}{c|}{\textbf{DCASE}} &
  \multicolumn{4}{c|}{\textbf{TUT}} \\ \hline
\textbf{} &
  \multicolumn{1}{c|}{\textbf{16b}} &
  \multicolumn{1}{c|}{\textbf{32b}} &
  \multicolumn{1}{c|}{\textbf{64b}} &
  \multicolumn{1}{c|}{\textbf{128b}} &
  \multicolumn{1}{l|}{\textbf{16b}} &
  \multicolumn{1}{l|}{\textbf{32b}} &
  \multicolumn{1}{l|}{\textbf{64b}} &
  \textbf{128b} &
  \multicolumn{1}{l|}{\textbf{16b}} &
  \multicolumn{1}{l|}{\textbf{32b}} &
  \multicolumn{1}{l|}{\textbf{64b}} &
  \textbf{128b} \\ \hline
AudioNet+TCL &
  \multicolumn{1}{l|}{0.325} &
  \multicolumn{1}{l|}{0.417} &
  \multicolumn{1}{l|}{0.515} &
  0.531 &
  \multicolumn{1}{l|}{0.437} &
  \multicolumn{1}{l|}{0.463} &
  \multicolumn{1}{l|}{0.581} &
  0.592 &
  \multicolumn{1}{l|}{0.403} &
  \multicolumn{1}{l|}{0.465} &
  \multicolumn{1}{l|}{0.574} &
  0.581 \\ \hline
AudioNet+WCL &
  \multicolumn{1}{l|}{0.443} &
  \multicolumn{1}{l|}{0.484} &
  \multicolumn{1}{l|}{0.711} &
  0.724 &
  \multicolumn{1}{l|}{0.697} &
  \multicolumn{1}{l|}{0.741} &
  \multicolumn{1}{l|}{0.854} &
  0.878 &
  \multicolumn{1}{l|}{0.582} &
  \multicolumn{1}{l|}{0.693} &
  \multicolumn{1}{l|}{0.802} &
  0.836 \\ \hline
AudioNet+WCL* &
  \multicolumn{1}{l|}{0.466} &
  \multicolumn{1}{l|}{0.502} &
  \multicolumn{1}{l|}{0.742} &
  0.756 &
  \multicolumn{1}{l|}{0.714} &
  \multicolumn{1}{l|}{0.771} &
  \multicolumn{1}{l|}{0.884} &
  0.892 &
  \multicolumn{1}{l|}{0.608} &
  \multicolumn{1}{l|}{0.724} &
  \multicolumn{1}{l|}{0.823} &
  0.848 \\ \hline
\end{tabular}
\end{table*}    

\subsection{Multi window fusion approach}
Determining the optimal sliding window length for sound analysis can be challenging. If the window is too short, it may fail to capture long-term variations in the signal. In contrast, if it is too long, it may obscure segmental boundaries between consecutive events and result in overlapping events within a frame. To address this issue, we utilized a multi-window feature fusion approach. This method involves extracting MFCC for three different window lengths: 4096, 11025, and 22050 samples. The features are then combined, forming a 3-channel input analogous to RGB channels in an image. Such a fusion approach ensures a comprehensive representation of the audio signal, significantly enhancing the model's performance in classifying diverse audio events.

\section{Results and discussion}

The performance of the proposed method is evaluated with three standard datasets and is reported for three variants of the AudioNet: AudioNet with traditional contrastive loss (AudioNet-TCL), AudioNet with proposed weighted contrastive loss (AudioNet-WCL), and AudioNet-WCL with multi-window fusion (AudioNet-WCL*) approach as discussed above. The mean average precision is reported for different hashcodes of bit lengths 16, 32, 64, and 128.

Table \ref{table@all} shows the mean average precision results for the three datasets for various bit lengths. Here, the mAP is evaluated for all retrieved items in the database. It is observed among the datasets that the performance of the model increases with bit length. However, there is a small improvement in the performance from 64 bits to 128 bits. The proposed method outperforms the conventional traditional loss and achieves a significantly higher mAP on DCASE and TUT acoustic compared to ESC-50.

\begin{figure*}[t!]

\subfloat[]{\includegraphics[width=2.2 in]{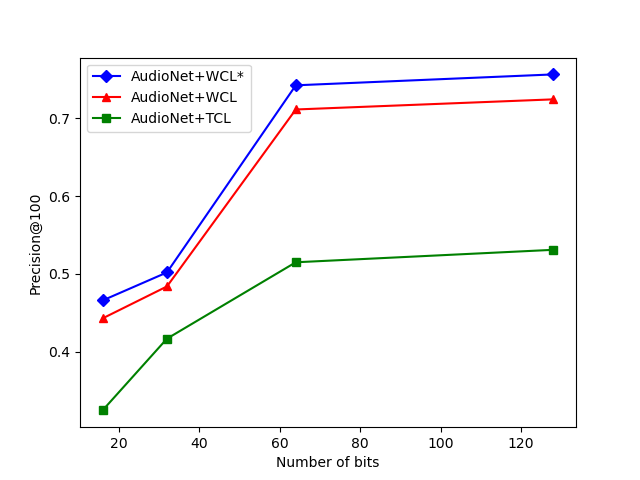}\label{e1}}%
\subfloat[]{\includegraphics[width=2.2 in]{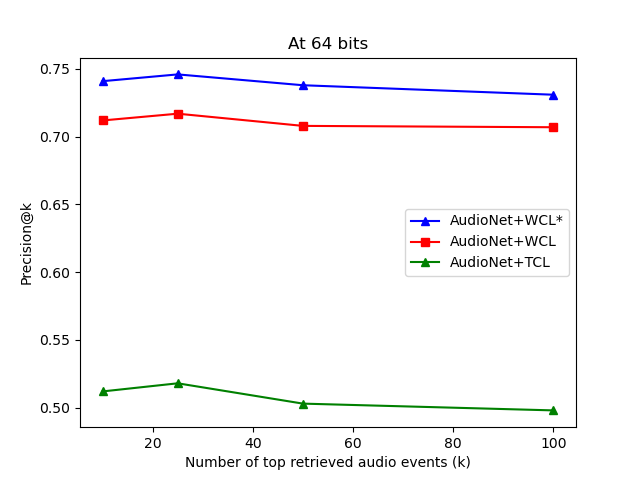}\label{e2}}%
\subfloat[]{\includegraphics[width=2.2 in]{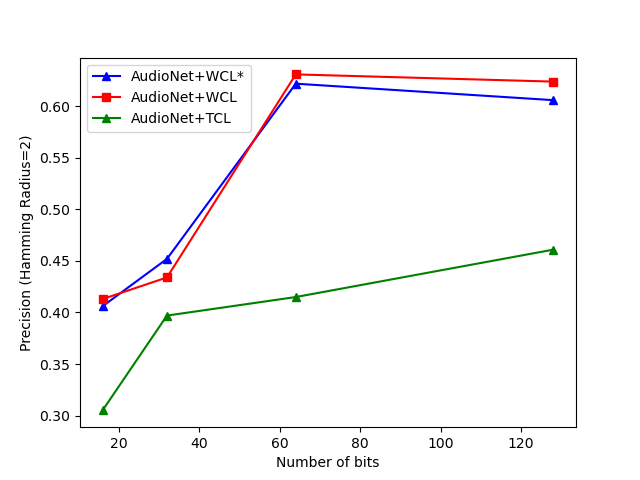}\label{e3}}%

\caption{ESC-50: a) Precision for top-100 for various bit length b) Precision at top k retrieved audio events with 64 bits c) Precision with hamming distance 2}
\label{map_esc50}
\end{figure*}

\begin{figure*}[t!]
\centering

\subfloat[]{\includegraphics[width=2.2 in]{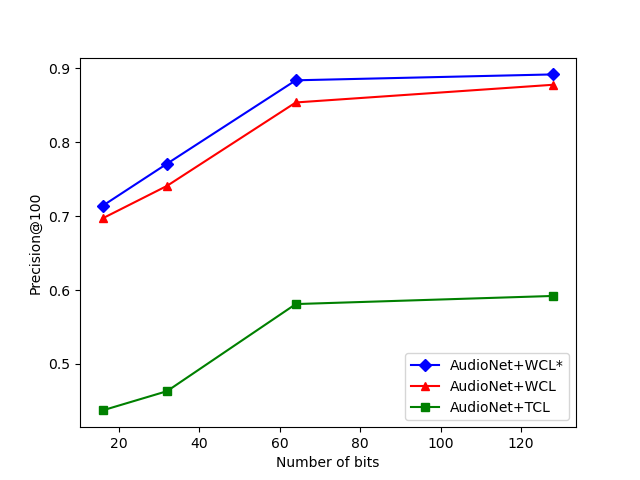}\label{d1}}%
\subfloat[]{\includegraphics[width=2.2 in]{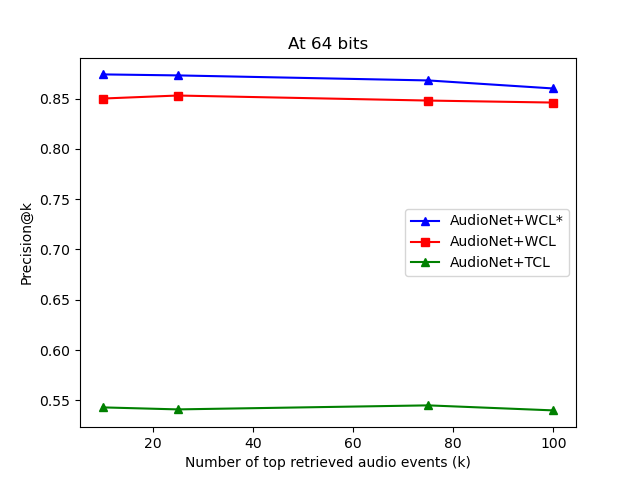}\label{d2}}%
\subfloat[]{\includegraphics[width=2.2 in]{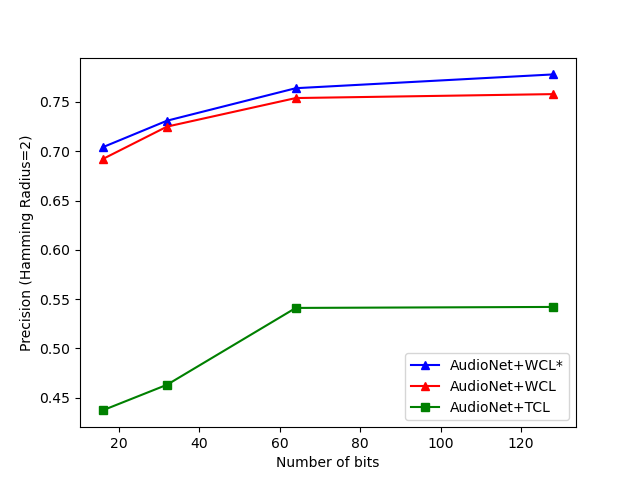}\label{d3}}%

\caption{DCASE: a) Precision for top-100 for various bit length b) Precision at top k retrieved audio events with 64 bits c) Precision with hamming distance 2}
\label{map_dcase}
\end{figure*}

\begin{figure*}[t!]
\centering

\subfloat[]{\includegraphics[width=2.2 in]{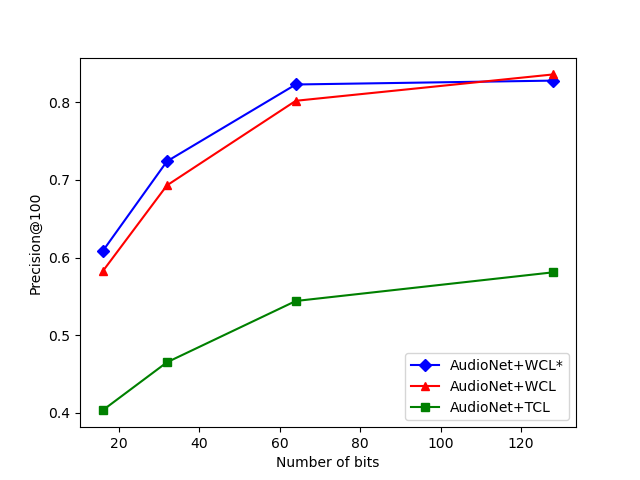}\label{t1}}%
\subfloat[]{\includegraphics[width=2.2 in]{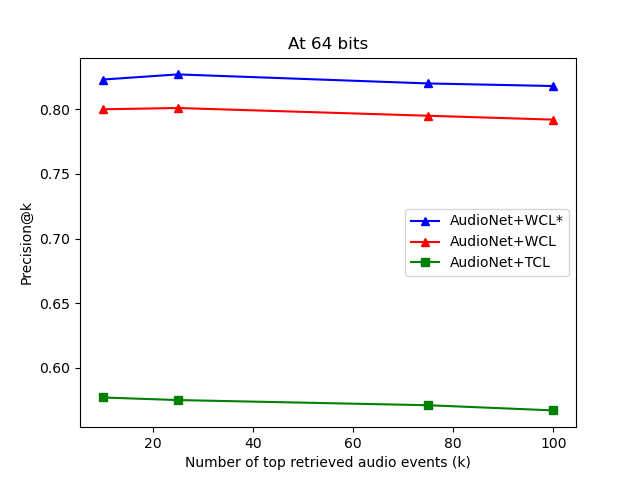}\label{t2}}%
\subfloat[]{\includegraphics[width=2.2 in]{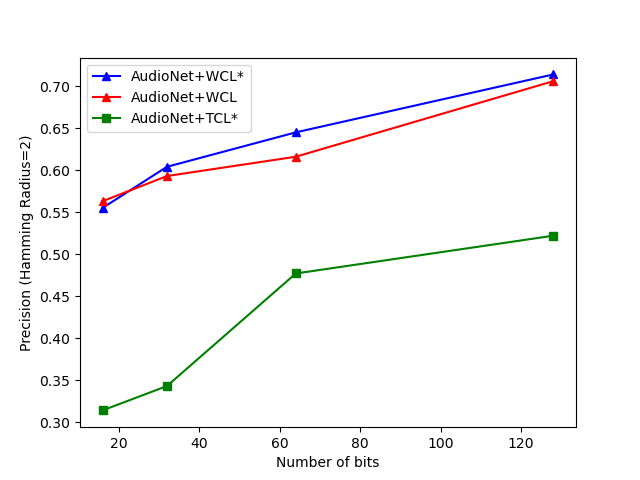}\label{t3}}%

\caption{TUT Acoustic Scene: a) Precision for top-100 for various bit lengths b) Precision at top k retrieved audio events with 64 bits c) Precision with hamming distance 2}
\label{map_tut}
\end{figure*}

The retrieval performance for the top 100 results for the three datasets is also evaluated and shown in Fig. \ref{e1}, Fig. \ref{d1}, and Fig. \ref{t1} and Table \ref{top100} summarizes the mAP for top-100 results. Here, the optimum bit length is observed to be 64-bit as there is no significant increase in mAP at 128-bit. The highest mAP of 74.2\%, 88.4\%, and 82.3\% is achieved by AudioNet+WCL* for ESC-50, DCASE, and TUT Acoustic Scenes datasets respectively. This validates that the multi-window feature input contributes to the increase in mAP. Compared to Audio DQN\cite{a10} where authors only reported the evaluation metric on the DCASE dataset, we achieved a 10\% boost in the mAP for 64-bit on the DCASE dataset for top 100 retrievals.

We also evaluated precision@k for various values of k at 64 bits to assess the robustness of the proposed approach. The metric precision@k represents performance when considering only the top k retrieved audios. Precision@k performance is shown in Fig. \ref{e2}, Fig. \ref{d2} and Fig. \ref{t2}. The precision for the top 100 retrieved audio is shown to have a flat curve, which suggests that the performance is stable. This shows the effectiveness of the suggested method in retrieving similar audio events.

Fig. \ref{e3}, Fig. \ref{d3} and Fig. \ref{t3} shows the precision with hamming radius 2. Achieving precise results within a hamming radius of 2 is crucial for the efficient retrieval of hash codes, as it ensures accurate results while balancing the trade-off between retrieval accuracy and false matches. It is a measure of the accuracy or correctness of retrieval when considering hashcodes that are within Hamming distance 2 from the audio query. It is seen that the performance of the proposed method with hamming radius 2 suffers a loss of precision but still retains a high precision value. 

Table \ref{percentage} shows the mean average precision for different percentages of training data for various bit lengths. It is observed that the proposed method performs worse when the training data is less. With AudioNet+WCL*, mAP for ESC-50, DCASE, and TUT acoustic scenes are found to be 48.3\%, 64.6\%, and 63.2\%, respectively, when the training data is 50\% of the database. The model performs relatively well with half the database. This illustrates that there is a scope for improvement in the overall method using less training data.

Table \ref{compare} presents the mean Average Precision (mAP) scores for various methods using 64-bit hashcodes across ESC-50, DCASE, and TUT datasets. The baseline methods encompass a wide range of classical unsupervised and pairwise similarity-based hashing techniques, including recent advancements in discrete hashing methods. The table also includes performances of pre-trained audio models such as PANNs, YAMNet, VGGish, and EfficientNetV2, all combined with WCL*, as well as ResNet models with Center Loss \cite{centerloss} and ArcFace \cite{arcface}. Our extensive experiments demonstrate that the proposed AudioNet model, particularly the variants AudioNet+WCL, and AudioNet+WCL*, significantly outperforms these baseline methods implemented for audio retrieval tasks. Additionally, the pre-trained PANNs model demonstrated notably good performance with our proposed method. The superior performance of our suggested methods, especially AudioNet+WCL* which achieves the highest mAP scores, establishes a strong foundation for future research into efficient retrieval of similar audio events using deep audio embeddings. These findings underscore the effectiveness of our approach in leveraging the power of deep learning for audio analysis and set a new benchmark in the field.

\begin{table*}[t]
\centering
\caption{Mean average precision $\uparrow$ for different percentages of the database for all retrieved audios}
\label{percentage}
\resizebox{\textwidth}{!}{%
\begin{tabular}{|c|c|cccc|cccc|cccc|}
\hline
\textbf{Method} &
  \textbf{\begin{tabular}[c]{@{}c@{}}\% of database\\ used for training\end{tabular}} &
  \multicolumn{4}{c|}{\textbf{ESC-50}} &
  \multicolumn{4}{c|}{\textbf{DCASE}} &
  \multicolumn{4}{c|}{\textbf{TUT}} \\ \hline
 &
   &
  \multicolumn{1}{c|}{\textbf{16b}} &
  \multicolumn{1}{c|}{\textbf{32b}} &
  \multicolumn{1}{c|}{\textbf{64b}} &
  \textbf{128b} &
  \multicolumn{1}{c|}{\textbf{16b}} &
  \multicolumn{1}{c|}{\textbf{32b}} &
  \multicolumn{1}{c|}{\textbf{64b}} &
  \textbf{128b} &
  \multicolumn{1}{c|}{\textbf{16b}} &
  \multicolumn{1}{c|}{\textbf{32b}} &
  \multicolumn{1}{c|}{\textbf{64b}} &
  \textbf{128b} \\ \hline
\multirow{4}{*}{\begin{tabular}[c]{@{}c@{}}AudioNet+\\ TCL\end{tabular}} &
  10\% &
  \multicolumn{1}{c|}{0.154} &
  \multicolumn{1}{c|}{0.214} &
  \multicolumn{1}{c|}{0.226} &
  0.224 &
  \multicolumn{1}{c|}{0.172} &
  \multicolumn{1}{c|}{0.246} &
  \multicolumn{1}{c|}{0.253} &
  0.259 &
  \multicolumn{1}{c|}{0.187} &
  \multicolumn{1}{c|}{0.253} &
  \multicolumn{1}{c|}{0.264} &
  0.279 \\ \cline{2-14} 
 &
  25\% &
  \multicolumn{1}{c|}{0.215} &
  \multicolumn{1}{c|}{0.292} &
  \multicolumn{1}{c|}{0.345} &
  0.356 &
  \multicolumn{1}{c|}{0.245} &
  \multicolumn{1}{c|}{0.274} &
  \multicolumn{1}{c|}{0.322} &
  0.337 &
  \multicolumn{1}{c|}{0.204} &
  \multicolumn{1}{c|}{0.305} &
  \multicolumn{1}{c|}{0.324} &
  0.337 \\ \cline{2-14} 
 &
  50\% &
  \multicolumn{1}{c|}{0.237} &
  \multicolumn{1}{c|}{0.336} &
  \multicolumn{1}{c|}{0.373} &
  0.380 &
  \multicolumn{1}{c|}{0.308} &
  \multicolumn{1}{c|}{0.325} &
  \multicolumn{1}{c|}{0.365} &
  0.377 &
  \multicolumn{1}{c|}{0.278} &
  \multicolumn{1}{c|}{0.322} &
  \multicolumn{1}{c|}{0.361} &
  0.378 \\ \cline{2-14} 
 &
  75\% &
  \multicolumn{1}{c|}{0.252} &
  \multicolumn{1}{c|}{0.347} &
  \multicolumn{1}{c|}{0.40} &
  0.418 &
  \multicolumn{1}{c|}{0.347} &
  \multicolumn{1}{c|}{0.403} &
  \multicolumn{1}{c|}{0.451} &
  0.467 &
  \multicolumn{1}{c|}{0.307} &
  \multicolumn{1}{c|}{0.366} &
  \multicolumn{1}{c|}{0.417} &
  0.428 \\ \hline
\multirow{4}{*}{\begin{tabular}[c]{@{}c@{}}AudioNet+\\ WCL\end{tabular}} &
  10\% &
  \multicolumn{1}{c|}{0.229} &
  \multicolumn{1}{c|}{0.255} &
  \multicolumn{1}{c|}{0.259} &
  0.264 &
  \multicolumn{1}{c|}{0.290} &
  \multicolumn{1}{c|}{0.354} &
  \multicolumn{1}{c|}{0.410} &
  0.426 &
  \multicolumn{1}{c|}{0.323} &
  \multicolumn{1}{c|}{0.366} &
  \multicolumn{1}{c|}{0.433} &
  0.445 \\ \cline{2-14} 
 &
  25\% &
  \multicolumn{1}{c|}{0.262} &
  \multicolumn{1}{c|}{0.307} &
  \multicolumn{1}{c|}{0.361} &
  0.378 &
  \multicolumn{1}{c|}{0.351} &
  \multicolumn{1}{c|}{0.418} &
  \multicolumn{1}{c|}{0.537} &
  0.553 &
  \multicolumn{1}{c|}{0.365} &
  \multicolumn{1}{c|}{0.433} &
  \multicolumn{1}{c|}{0.527} &
  0.534 \\ \cline{2-14} 
 &
  50\% &
  \multicolumn{1}{c|}{0.282} &
  \multicolumn{1}{c|}{0.436} &
  \multicolumn{1}{c|}{0.455} &
  0.468 &
  \multicolumn{1}{c|}{0.463} &
  \multicolumn{1}{c|}{0.502} &
  \multicolumn{1}{c|}{0.623} &
  0.634 &
  \multicolumn{1}{c|}{0.399} &
  \multicolumn{1}{c|}{0.498} &
  \multicolumn{1}{c|}{0.570} &
  0.583 \\ \cline{2-14} 
 &
  75\% &
  \multicolumn{1}{c|}{0.303} &
  \multicolumn{1}{c|}{0.484} &
  \multicolumn{1}{c|}{0.561} &
  0.574 &
  \multicolumn{1}{c|}{0.507} &
  \multicolumn{1}{c|}{0.641} &
  \multicolumn{1}{c|}{0.704} &
  0.713 &
  \multicolumn{1}{c|}{0.439} &
  \multicolumn{1}{c|}{0.554} &
  \multicolumn{1}{c|}{0.644} &
  0.653 \\ \hline
\multirow{4}{*}{\begin{tabular}[c]{@{}c@{}}AudioNet+\\ WCL*\end{tabular}} &
  10\% &
  \multicolumn{1}{c|}{0.244} &
  \multicolumn{1}{c|}{0.267} &
  \multicolumn{1}{c|}{0.271} &
  0.278 &
  \multicolumn{1}{c|}{0.317} &
  \multicolumn{1}{c|}{0.365} &
  \multicolumn{1}{c|}{0.421} &
  0.428 &
  \multicolumn{1}{c|}{0.344} &
  \multicolumn{1}{c|}{0.386} &
  \multicolumn{1}{c|}{0.452} &
  0.459 \\ \cline{2-14} 
 &
  25\% &
  \multicolumn{1}{c|}{0.275} &
  \multicolumn{1}{c|}{0.311} &
  \multicolumn{1}{c|}{0.372} &
  0.379 &
  \multicolumn{1}{c|}{0.377} &
  \multicolumn{1}{c|}{0.407} &
  \multicolumn{1}{c|}{0.550} &
  0.578 &
  \multicolumn{1}{c|}{0.372} &
  \multicolumn{1}{c|}{0.414} &
  \multicolumn{1}{c|}{0.559} &
  0.565 \\ \cline{2-14} 
 &
  50\% &
  \multicolumn{1}{c|}{0.306} &
  \multicolumn{1}{c|}{0.453} &
  \multicolumn{1}{c|}{0.483} &
  0.492 &
  \multicolumn{1}{c|}{0.457} &
  \multicolumn{1}{c|}{0.542} &
  \multicolumn{1}{c|}{0.646} &
  0.665 &
  \multicolumn{1}{c|}{0.418} &
  \multicolumn{1}{c|}{0.522} &
  \multicolumn{1}{c|}{0.632} &
  0.651 \\ \cline{2-14} 
 &
  75\% &
  \multicolumn{1}{c|}{0.344} &
  \multicolumn{1}{c|}{0.482} &
  \multicolumn{1}{c|}{0.552} &
  0.566 &
  \multicolumn{1}{c|}{0.538} &
  \multicolumn{1}{c|}{0.651} &
  \multicolumn{1}{c|}{0.718} &
  0.727 &
  \multicolumn{1}{c|}{0.442} &
  \multicolumn{1}{c|}{0.567} &
  \multicolumn{1}{c|}{0.707} &
  0.718 \\ \hline
\end{tabular}}
\end{table*}

\begin{table}[h]
\centering
\caption{Mean Average Precision $\uparrow$ for all retrieved audio events for different methods at 64 bit}
\label{compare}
\begin{tabular}{cccc}
\hline
\textbf{Method} & \multicolumn{1}{l}{\textbf{ESC-50}} & \multicolumn{1}{l}{\textbf{DCASE}} & \multicolumn{1}{l}{\textbf{TUT}} \\ \hline
Locality Sensitive Hashing \cite{para3a}        & 0.027          & 0.064          & 0.034          \\
Spectral Hashing \cite{d27}                  & 0.161          & 0.126          & 0.193          \\
Iterative Quantization \cite{d26}            & 0.258          & 0.152          & 0.275          \\
Anchor Graph Hashing \cite{agh}              & 0.276          & 0.141          & 0.251          \\
Product Quantization \cite{pq}              & 0.231          & 0.144          & 0.246          \\
DSDH \cite{dsdh}                              & 0.491          & 0.574          & 0.550          \\
GreedyHash \cite{greedy}                        & 0.571          & 0.532          & 0.587          \\
DHN \cite{dhn}                               & 0.550          & 0.503          & 0.594          \\
DPSH \cite{dpsh}                              & 0.591          & 0.486          & 0.5317         \\
AudioDQN \cite{a10}                          & -              & ~70            & -              \\
Boosted Locality Sensitive Hashing \cite{kim2022boosted} & 0.483          & 0.527          & 0.562          \\
OrthoHash \cite{hoe2021one}                         & 0.486          & 0.522          & 0.577          \\
PANNs+WCL* \cite{kong2020panns}                        & 0.594          & 0.653          & 0.542          \\
YAMNet+WCL* \cite{howard2017mobilenets}                       & 0.428          & 0.508          & 0.395          \\
VGGish+WCL* \cite{m2}                       & 0.396          & 0.434          & 0.376          \\
EfficientNetV2+WCL* \cite{nguyen2023fruit}               & 0.323          & 0.347          & 0.307          \\
ResNet \cite{m4} + Center Loss \cite{centerloss}               & 0.292          & 0.366          & 0.337          \\
ResNet \cite{m4} + ArcFace \cite{arcface}                  & 0.194          & 0.231          & 0.205          \\
\textit{AudioNet+TCL}              & 0.439          & 0.53           & 0.487          \\
\textit{AudioNet+WCL}              & \textbf{0.671} & \textbf{0.784} & \textbf{0.754} \\
\textit{AudioNet+WCL*}             & \textbf{0.692} & \textbf{0.808} & \textbf{0.778} \\ \hline
\end{tabular}
\end{table}

Table \ref{time} presents the retrieval times for various methods using the DCASE dataset for all retrieved items. A 10-second audio query was selected to compare retrieval times across different methods. The model parameters are consistent with those used to calculate the mAP in Table \ref{compare}. The table shows retrieval times and mAP @ 64 bits for various methods. Methods like Locality Sensitive Hashing and Spectral Hashing have high retrieval times (45.32-59.20s) and low mAP scores (0.064-0.152), indicating inefficient retrieval performance. In contrast, methods like DSDH, Greedy Hash, DHN, and DPSH show moderate retrieval times (21.89-33.12s) and better mAP scores (0.486-0.574). The proposed method, AudioNet+WCL*, achieves the best performance with the lowest retrieval times (13.37 to 16.70 seconds) and the highest mAP score (0.808), highlighting its superior retrieval accuracy and efficiency. 


Two evaluation protocols for supervised hashing are studied in recent computer vision literature \cite{z1}, a supervised retrieval protocol, in which the classes of the queries and the database are identical, and a zero-shot retrieval protocol, in which the classes of the queries and the database are different. The performance of some supervised hashing methods varies greatly depending on the protocol used. One may perform well on one protocol and poorly on another. Table \ref{zero} illustrates how AudioNet-WCL performs for the zero-shot protocol and demonstrates how it performs well with other protocols. 

\begin{figure}[t]
\centering
\subfloat[]{\includegraphics[width=3.5 in]{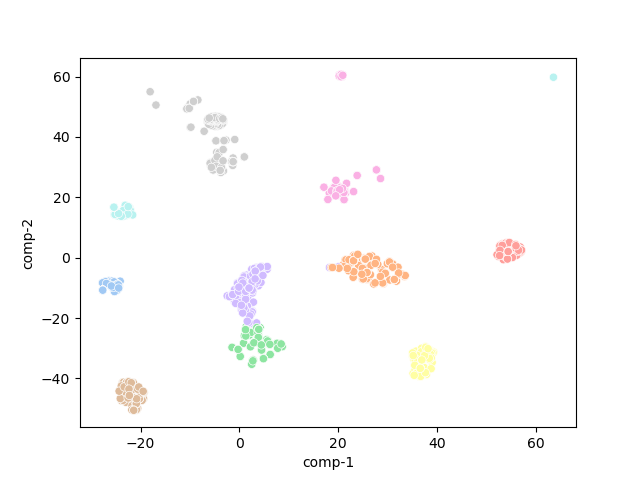}}
\caption{t-SNE visualization of hash codes for DCASE Task-2}
\label{tsne}
\end{figure}

Fig. \ref{tsne} shows the t-SNE visualization of hash codes generated by AudioNet on DCASE. We select 10 random classes for visualization. AudioNet can generate discriminative hash codes, which in turn enables improved performance in similarity retrieval tasks. By learning and generating hash codes that effectively capture the discriminative features of audio data, AudioNet can enhance the accuracy and efficiency of similarity-based retrieval tasks, such as audio matching, audio recognition, and audio retrieval, leading to more effective and reliable results in various audio-related applications.

\begin{table}[t]
\centering
\caption{Comparison of Retrieval Times Across Various Methods Using a 10-Second Audio Query from the DCASE Task-2 Dataset}
\label{time}
\begin{tabular}{@{}lcccc@{}}
\toprule
\multicolumn{1}{c}{\textbf{Method}} &
  \multicolumn{3}{c}{\textbf{Retrieval time (s)}} &
  \multicolumn{1}{l}{} \\ \midrule
 &
  \multicolumn{1}{l}{\textbf{16b}} &
  \multicolumn{1}{l}{\textbf{32b}} &
  \multicolumn{1}{l}{\textbf{64b}} &
  \multicolumn{1}{l}{\textbf{mAP @ 64b}} \\ \cmidrule(l){2-5} 
Locality Sensitive Hashing \cite{para3a} & 45.32 & 50.45 & 55.67 & 0.064 \\
Spectral Hashing \cite{d27}          & 46.78 & 52.34 & 57.89 & 0.126 \\
Iterative Quantization \cite{d26}    & 47.12 & 53.23 & 58.12 & 0.152 \\
Anchor Graph Hashing \cite{agh}      & 47.5  & 54.89 & 59.2  & 0.141 \\
Product Quantization \cite{pq}      & 48.56 & 53.23 & 58.12 & 0.144 \\
DSDH \cite{dsdh}                      & 22.45 & 25.32 & 29.45 & 0.574 \\
GreedyHash \cite{greedy}                & 26.78 & 29.45 & 33.12 & 0.532 \\
DHN \cite{dhn}                       & 24.56 & 27.32 & 31.23 & 0.503 \\
DPSH \cite{dpsh}                      & 21.89 & 23.45 & 27.12 & 0.486 \\
AudioNet+WCL*              & 13.37 & 14.36 & 16.7  & 0.808 \\ \bottomrule
\end{tabular}
\end{table}

\begin{table}[t]
\centering
\caption{Mean average precision $\uparrow$ with the zero-shot protocol for DCASE}
\label{zero}
\begin{tabular}{@{}lllll@{}}
\toprule
\textbf{Method} &
  \multicolumn{1}{c}{\textbf{16b}} &
  \multicolumn{1}{c}{\textbf{32b}} &
  \multicolumn{1}{c}{\textbf{64b}} &
  \multicolumn{1}{c}{\textbf{128b}} \\ \midrule
AudioNet+TCL  & 0.225 & 0.304 & 0.375 & 0.392 \\
AudioNet+WCL  & 0.432 & 0.545 & 0.637 & 0.652 \\
AudioNet+WCL* & 0.488 & 0.583 & 0.685 & 0.694 \\ \bottomrule
\end{tabular}
\end{table}

\subsection{Ablation Study}
We conducted an ablation study to scrutinize the impact of different components of our loss function on the performance of the proposed AudioNet. This study is essential for understanding how variations in the loss function parameters influence the model's effectiveness in audio event retrieval tasks.

\begin{table*}[]
\centering
\caption{Comparative Retrieval Performance Analysis of various Datasets at 64-Bit}
\label{dataset}
\begin{tabular}{@{}llccccc@{}}
\toprule
\multicolumn{2}{c}{Dataset} &
  \begin{tabular}[c]{@{}c@{}}Total \\ Duration\\ (min)\end{tabular} &
  \begin{tabular}[c]{@{}c@{}}Total Sample \\ Count\end{tabular} &
  Classes &
  \begin{tabular}[c]{@{}c@{}}mAP for all\\ retrieved @ 64b\end{tabular} &
  \multicolumn{1}{l}{\begin{tabular}[c]{@{}l@{}}mAP for Top-100 \\ retrieved @ 64b\end{tabular}} \\ \midrule
\multicolumn{1}{c}{\multirow{3}{*}{Balanced}} & ESC-50                         & 2.8 hr   & 2000   & 50  & 0.692 & 0.742 \\
\multicolumn{1}{c}{}                          & TUT Acoustic Scenes 2017       & 13 hr    & 4680   & 15  & 0.778 & 0.823 \\
\multicolumn{1}{c}{}                          & TAU Urban Acoustic Scenes 2020 & 64 hr    & 23030  & 10  & 0.821 & 0.873 \\ \midrule
Imbalanced                                    & 2018 DCASE Task-2              & 23 hr    & 9500   & 41  & 0.808 & 0.884 \\
                                              & FSD50K                         & 108.3 hr & 51197  & 200 & 0.654 & 0.713 \\
                                              & FSDnoisy18k                    & 42.5     & 18533  & 20  & 0.744 & 0.821 \\
                                              & 2018 DCASE Task-5              & 202 hr   & 72984  & 9   & 0.832 & 0.906 \\
                                              & EPIC-SOUNDS                    & 100 hr   & 78366  & 44  & 0.678 & 0.752 \\
                                              & SPASS                          & 69.4 hr  & 351340 & 28  & 0.811 & 0.872 \\ \bottomrule
\end{tabular}
\end{table*}

\begin{table*}[t]
\centering
\caption{Comparative Analysis of Model Performance Using Individual Components of Loss Function}
\label{ablation2}
\begin{tabular}{|l|cccc|cccc|cccc|}
\hline
\multicolumn{1}{|c|}{\textbf{Method}} &
  \multicolumn{4}{c|}{\textbf{ESC-50}} &
  \multicolumn{4}{c|}{\textbf{DCASE}} &
  \multicolumn{4}{c|}{\textbf{TUT}} \\ \hline
\multicolumn{1}{|c|}{} &
  \multicolumn{1}{c|}{\textbf{16b}} &
  \multicolumn{1}{c|}{\textbf{32b}} &
  \multicolumn{1}{c|}{\textbf{64b}} &
  \textbf{128b} &
  \multicolumn{1}{c|}{\textbf{16b}} &
  \multicolumn{1}{c|}{\textbf{32b}} &
  \multicolumn{1}{c|}{\textbf{64b}} &
  \textbf{128b} &
  \multicolumn{1}{c|}{\textbf{16b}} &
  \multicolumn{1}{c|}{\textbf{32b}} &
  \multicolumn{1}{c|}{\textbf{64b}} &
  \textbf{128b} \\ \hline
AudioNet (WCL*) &
  \multicolumn{1}{c|}{0.376} &
  \multicolumn{1}{c|}{0.632} &
  \multicolumn{1}{c|}{0.692} &
  0.698 &
  \multicolumn{1}{c|}{0.638} &
  \multicolumn{1}{c|}{0.751} &
  \multicolumn{1}{c|}{0.808} &
  0.817 &
  \multicolumn{1}{c|}{0.512} &
  \multicolumn{1}{c|}{0.677} &
  \multicolumn{1}{c|}{0.773} &
  0.780 \\ \hline
AudioNet (WCL*) + $C_{loss}$ &
  \multicolumn{1}{c|}{0.346} &
  \multicolumn{1}{c|}{0.584} &
  \multicolumn{1}{c|}{0.640} &
  0.645 &
  \multicolumn{1}{c|}{0.589} &
  \multicolumn{1}{c|}{0.694} &
  \multicolumn{1}{c|}{0.747} &
  0.755 &
  \multicolumn{1}{c|}{0.473} &
  \multicolumn{1}{c|}{0.626} &
  \multicolumn{1}{c|}{0.715} &
  0.722 \\ \hline
AudioNet (WCL*) + $\mathcal{L}_{p}$ &
  \multicolumn{1}{c|}{0.278} &
  \multicolumn{1}{c|}{0.474} &
  \multicolumn{1}{c|}{0.518} &
  0.522 &
  \multicolumn{1}{c|}{0.474} &
  \multicolumn{1}{c|}{0.563} &
  \multicolumn{1}{c|}{0.605} &
  0.611 &
  \multicolumn{1}{c|}{0.379} &
  \multicolumn{1}{c|}{0.504} &
  \multicolumn{1}{c|}{0.576} &
  0.582 \\ \hline
\end{tabular}
\end{table*}


\begin{table}[t]
\centering
\caption{Performance based on different combinations of $\lambda$ and $\beta$}
\label{ablation1}
\begin{tabular}{@{}cllccc@{}}
\toprule
\multicolumn{1}{l}{\textbf{Method}} &
  \textbf{$\lambda$} &
  \textbf{$\beta$} &
  \multicolumn{1}{l}{\textbf{ESC-50}} &
  \multicolumn{1}{l}{\textbf{DCASE}} &
  \multicolumn{1}{l}{\textbf{TUT}} \\ \midrule
\multirow{5}{*}{AudioNet+WCL*} & 0.1 & 0.1 & 0.358 & 0.592 & 0.514 \\
                               & 0.1 & 0.5 & 0.485 & 0.572 & 0.587 \\
                               & 0.5 & 0.1 & 0.631 & 0.753 & 0.694 \\
                               & 0.5 & 0.5 & 0.675 & 0.783 & 0.753 \\ 
                               & 0.7 & 0.3 & 0.687 & 0.792 & 0.768 \\
                               \bottomrule
\end{tabular}
\end{table}

In the first part of our ablation study (Table \ref{ablation1}), we focused on adjusting the weights, $\lambda$ and $\beta$, in the loss function to see how these changes would affect performance on three datasets: ESC-50, DCASE, and TUT, using 64-bit hashcodes. The goal is to find the right balance between two parts of the loss function: the contrastive loss ($C_{loss}$) and the pairwise loss ($\mathcal{L}_{p}$). The results showed that setting both $\lambda$ and $\beta$ to 0.5 provided the best balance, improving the model’s ability to retrieve audio events across all datasets. This suggested that an equal contribution from both loss components is beneficial. However, aiming to further optimize the model, we fine-tuned the weights to $\lambda$ = 0.7 and $\beta$ = 0.3. This decision is based on the idea that slightly favoring the contrastive loss might improve the model’s performance. However, it is important to highlight that the model demonstrated a notable degree of robustness to minor changes in $\lambda$ and $\beta$. This indicates that the performance of the model is not significantly impacted by marginal adjustments to these parameters, suggesting an inherent stability in the model's architecture against such variations.

In the second part of our study (Table \ref{ablation2}), we evaluated the performance of the model using each component of the loss function independently. This included three scenarios: using only the weighted contrastive loss (WCL*), using WCL* combined with $C_{loss}$, and using WCL* combined with $\mathcal{L}_{p}$. The performance is measured across the ESC-50, DCASE, and TUT datasets and for different binary hash lengths (16b, 32b, 64b, 128b).

The findings from Table \ref{ablation2} indicated that using both the components ($C_{loss}$ + $\mathcal{L}{p}$) led to superior performance compared to integrating it with either $C_{loss}$ or $\mathcal{L}{p}$ alone. This suggests that while both $C_{loss}$ and $\mathcal{L}_{p}$ contribute to performance improvement, using both the components plays a more crucial role in the model's ability to effectively retrieve similar audio events. For Table \ref{ablation2}, the average margin percentage decreases are approximately 7.54\% when comparing the performance of using only WCL* against WCL* combined with $C_{loss}$, and about 25.27\% when comparing WCL* alone against its combination with $\mathcal{L}_{p}$.

The ablation study provides valuable insights into the functioning and optimization of our proposed loss function. It is evident that a balanced approach to combining $C_{loss}$ and $\mathcal{L}_{p}$ is beneficial, as seen in the enhanced performance with equal weights for both components. Furthermore, the dominant role of WCL* in the model's performance underscores its importance in our loss function formulation. This analysis not only validates our approach but also guides future efforts in fine-tuning loss function components for audio event retrieval tasks.

\section{Conclusion}
In conclusion, this study introduces AudioNet, a supervised deep hashing technique designed for the retrieval of similar audio events. The approach addresses the need for efficient retrieval by employing deep audio embeddings and a novel loss function that combines weighted contrastive and pairwise losses. This method significantly outperforms existing hashing techniques in systematic evaluations conducted on three audio events benchmark datasets. The results confirm AudioNet's effectiveness and suggest its potential as a benchmark for future research in audio event retrieval.

Potential future work in the field of deep hashing for audio event retrieval could involve several areas of improvement. These could include further enhancements to deep learning models, incorporating multi-modal information for more comprehensive analysis, and exploring unsupervised or self-supervised learning approaches for improved performance. Additionally, investigating novel similarity measures to better capture audio event characteristics and evaluating real-world datasets to validate model effectiveness are important areas to consider. While we currently use MFCCs for their lower computational demand, we recognize that log-mel spectrograms offer richer frequency information and could better capture audio features. Future research may explore their integration to enhance performance across diverse audio conditions.

\section{Acknowledgement}

This work has been supported by the research grant (PB/EE/2021128-B) from Prasar Bharati, India’s National Broadcasting Company.

\bibliographystyle{IEEEtran} 
\bibliography{reference}


\newpage
\begin{IEEEbiography}[{\includegraphics[width=1in,height=1.25in, clip,keepaspectratio]{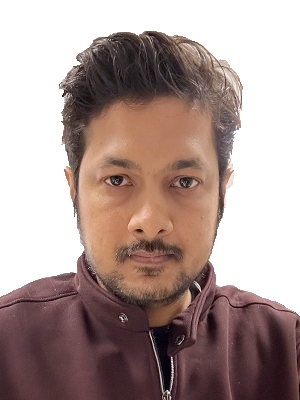}}]
{Sagar Dutta} received his Ph.D. degree in Electronics and Communication Engineering from the National Institute of Technology Silchar, India, in 2022. He was a postdoctoral researcher at IIT Kanpur and is currently a postdoctoral researcher at the RITMO Centre for Interdisciplinary Studies in Rhythm, Time, and Motion, Department of Musicology, University of Oslo. His research interests include machine learning, audio signal processing, music information retrieval, and motion capture.
\end{IEEEbiography}

\vskip -2\baselineskip plus -1fil

\begin{IEEEbiography}[{\includegraphics[width=1in,height=1.25in, clip,keepaspectratio]{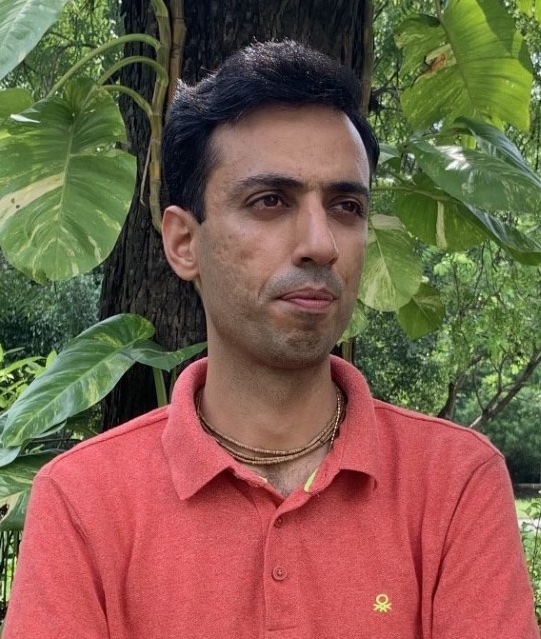}}]
{Vipul Arora} received his B.Tech. and Ph.D. degrees in Electrical Engineering from the Indian Institute of Technology (IIT) Kanpur, India, in 2009 and 2015, respectively. He was a Postdoctoral Researcher at the University of Oxford and a Research Scientist at Amazon Alexa, Boston, MA, USA. He is currently working as an Associate Professor with the Department of Electrical Engineering at IIT Kanpur. His research interests include machine learning, audio processing, machine learning for physics, and time series analysis.
\end{IEEEbiography}

\end{document}